# Time Efficient Training of Progressive Generative Adversarial Network using Depthwise Separable Convolution and Super Resolution Generative Adversarial Network

Atharva Karwande, Pranesh Kulkarni, Tejas Kolhe, Akshay Joshi, Soham Kamble

*Abstract*— Generative Adversarial Networks have been employed successfully to generate high-resolution augmented images of size $1024^2$. Although the augmented images generated are unprecedented, training time of the model is exceptionally high. Conventional GAN requires training of both Discriminator as well as the Generator. In Progressive GAN, which is the current state-of-the-art GAN for image augmentation, instead of training the GAN all at once, a new concept of progressing growing of Discriminator and Generator simultaneously, was proposed. Although the lower stages such as 4x4 and 8x8 train rather quickly, the later stages consume tremendous amount of time which could take days to finish the model training. In our paper we propose a novel pipeline that combines Progressive GAN with slight modifications and Super Resolution GAN. Super Resolution GAN up samples low resolution images to high resolution images which can prove to be a useful resource to reduce the training time exponentially.

## I. Introduction

The introduction of GAN by Ian Goodfellow et al, has provided an alternate dimension to the data augmentation task [1]. Generating high quality augmented images has been a priority ever since. While traditional techniques as proposed in [2-5] and autoregressive models [6-8] produces good images, they have not been able to keep up with the unparalleled images generated using state-of-the-art GANs.

A traditional GAN consists of two parts: Discriminator and a Generator. The generator produces an image from a latent space whose distribution must preferably be identical to the training distribution. To avoid distinguishing of these images using a mathematical function, a Discriminator is employed. It is trained to assess, whether the image passed through it is real (i.e., from the training set) or fake (i.e., generated using Generator). A gradient is employed to train the both the networks in the right direction. Once the Discriminator is trained, the Generator is of main interest.

Different loss functions have been used to measure the distance between the training distribution and the generated distribution such as Jenson-Shannon divergence [1], least square [9], and Wasserstein distance [10-11]. But one of the most effective metrics used is by the Progressive GAN, which is the improved Wasserstein distance [12].

One of the most challenging tasks is to generate high resolution images because, in high resolution, it is easy for the discriminator to distinguish between the fake and the real as it amplifies the gradient problem [13]. To tackle this problem, a new design was proposed in which use the strategy of growing the Discriminator and the Generator simultaneously and progressively starting from low resolution of 4x4, 8x8 onwards [12]. This greatly improves the training stability in the GAN. Although [14] proposes a new strategy to make the training robust in progressive GAN, the main aim of our paper is to provide a new pipeline which employees a state-of-the-art GAN and an algorithm that up samples the image effectively to reduce the training time drastically.

For up sampling, we use Super Resolution GAN [15] whose output is better than any other up sampling technique. Now, the degree of perceived variation in the output image has been in discussion. Various method employed for evaluating the images include the inception score [16], multiscale structural similarity [17] and birthday paradox [18]. But a new metric, sliced Wasserstein distance, has been proposed in [12] which captures the similarity in both appearance and spatial resolution.

We evaluate our pipeline using CelebA Dataset which is considered as benchmark for GANs. The evaluation metric used is sliced Wasserstein distance and multiscale structural similarity.

## II. Background and Related Work

To optimize or evaluate any algorithm, all the related work must be accounted for. Tero Karras et al. [12] has proposed a novel training process for generative adversarial networks for generating high resolution images with better stability and better image quality. In the proposed training methodology, the size of generator and discriminator increases progressively after certain set of epochs and resolution of images also increases progressively after each set of epochs. This process decreases the training time and increases stability of training process. The quality of generated images was outstanding compared with previously proposed GANs. Generally generative adversarial networks consist of two networks generator and discriminator. Generator generated the images from random noise (latent) and discriminator differentiates between images in the dataset (real images) and generated outputs by generator network (fake images). The network should be trained in such a way that generator should generate more realistic images i.e., the loss between training data and generated data should be minimum and discriminator should distinguish more precisely between real images and fake images. The author stated that in the method of progressively growing architecture, model discovers large scale features in the initial sets of epochs and after some certain epochs, it learns to find finer scale details in the sample. Author faded the new layers smoothly while doubling the architecture to avoid sudden shocks to well-trained model parameters. The key feature of the proposed method is to increase variation using minibatch standard deviation to encounter the problem of the tendency of GANS of capturing only subset of variation found in training data. Another key feature is equalized learning rate. In this paper, author proposed that by scaling the weights at each layer with a constant, such that the updated weight is scaled to w/c where w is original weight and c is constant at each layer to keep the weights in the model in similar scale while training process. For evaluation of GANs or to compare the results between GANs, author proposed a new evaluation matrix called Multi-Scale Statistical Similarity. Author implemented this methodology to generate images from some state-of-the-art datasets like CELEBA-HQ, LASUN bedroom, CIFAR10 datasets. The overall quality of generated images was very high. Zhou Wang et. al. [17] proposed a better method to evaluate and compare quality of images than mean squared error (MSE) i.e., structural similarity (SSIM) index. The system separates similarity measurement into three comparisons: luminance, contrast, and structure. However, the SSIM index algorithm is a single-scale approach. This can contribute to drawback because right scale depends on viewing conditions. Here authors propose a multi-scale structural similarity approach for image quality assessment, which provides more flexibility than single-scale approach in incorporating the variations of image resolution and viewing conditions. I. Deshpande et. al. [19] another method to evaluate the quality of image by calculating and comparing the Wasserstein distance. A small Wasserstein distance indicates that the distribution of the patches is similar, meaning that the training images and generator samples appear similar in appearance. Authors implemented 'sliced Wasserstein Distances' for experimentation purpose and found out that, the sliced Wasserstein distance is a good measure for distance between two distributions, considering both, the sample quality, and the sample diversity. C. Ledig et. al. [15] proposed the first framework capable of inferring photo-realistic natural image for 4 X upscaling factors. To achieve this, they propose a perceptual loss function which consist of an adversarial loss and a content loss. The adversarial loss pushes solution the natural image manifold using discriminator network. Discriminator network used to train to differentiate between the super-resolved images and original photo-realistic images. Content loss motivated by perceptual similarity instead of similarity in pixel space. An extensive mean-opinion-score (MOS) test shows hugely significant gains in perceptual quality using SRGAN. It is also seen that the MOS scores obtained with SRGAN are closer to the original high-resolution images than to those obtained with any state-of-the-art method.

## III. DATASET

The dataset used for evaluation of our pipeline is CelebA. Down sampling of the dataset was carried out using various technique. The best technique was selected using the time taken to down sample and the visual quality of the down sampled image. Table 1 summarizes the results.

| SR. No. | Algorithm | Time/Image (in s) | Visual Quality |
|---|---|---|---|
| 1. | Bicubic | 0.000977796 | 4 |
| 2. | Bilinear | 0.000912458 | 3 |
| 3. | Lanczos | 0.001175824 | 5 |
| 4. | Hamming | 0.000929764 | 3 |
| 5. | Nearest Neighbor | 0.000772037 | 1 |
| 6. | Box | 0.000868146 | 2 |

Table 1. Analysis of down sampling techniques

Upon closer viewing, Bicubic down sampling appears to be the best choice. We support our analysis by stating the fact that this algorithm provides a perfect tradeoff between quality and time to down sample. With nearest neighbor being the fastest with the lowest quality, and the Lanczos being the best quality but the slowest. As we are trying to reduce the training time using our pipeline, every second of time saved is worth it.

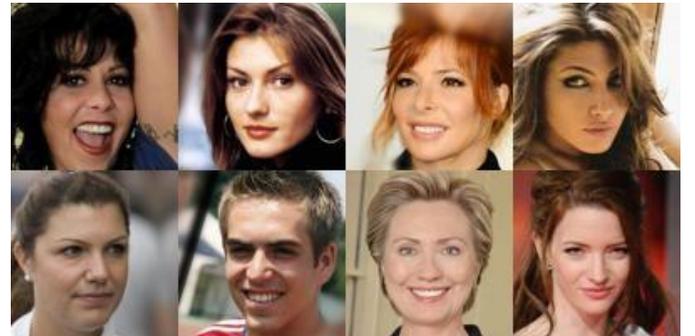

Fig. 1. Original CelebA Dataset of 64 x 64

Fig. 1. is the original dataset down sampled from 1024x1024 to 64x64.

## IV. DEPTHWISE SEPARABLE PROGRESSIVE GAN

The effectiveness of training a Convolutional Neural Network in terms of time saving has been portrayed in the MobileNets, Xception, etc. [20-21]. Replacing all the Vanilla Convolution with Depthwise Separable Convolution in Progressive GAN, we save enormous time without affecting the quality of the images much. Depthwise Separable Convolution can be compared with vanilla convolution with regards to the number of multiplications.

If the input dimension is $D_F \times D_F \times M$ and let the kernel size be $D_K \times D_K$ and let there be N kernels. The output size after convolution is $D_G \times D_G \times N$.
For Vanilla Convolution:

No. of Multiplications = $N \times D_K \times D_K \times D_F \times D_F \times M$
----------------- (1)

For Depthwise Separable Convolution:

No. of Multiplications = $M \times D_G \times D_G \times ((D_K \times D_K) + N))$
----------------- (2)

Dividing the equation (2) by (1), we have,
Depthwise Separable by Vanilla Convolution =

$(1/N) + (1/(D_K \times D_K))$
----------------- (3)

Equation (3) can be realized evidently by observing the training time graphs of Progressive GAN using Depthwise Separable Convolution and Vanilla Convolution for the resolution 64 x 64 in Fig. 2.

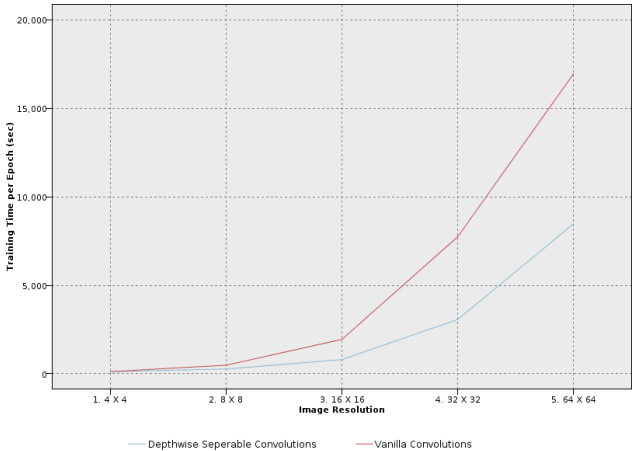

Fig. 2. Training time for Progressive GAN using Depthwise Separable Convolution (blue) and Vanilla Convolution (red)

Due to the limited training resources, training the Progressive GAN was possible up till the resolution of 64 x 64. It is evident that the Depthwise Separable Convolution speeds up the training by a factor of 2x. This is a tremendous decrease in the training time for a model which takes days to train.
The above model was trained on Google Collaboratory which is a free to use platform with GPU computing.

## V. UP SAMPLING BY SRGAN

The way progressive GAN works is by increasing the resolution by adding new layer. 4 x 4, 8 x 8, 16 x 16, and so on. As the training resolution increases, the training time increases exponentially. Even if we could eliminate one resolution training, then we can save days of training time. As the resolution increases, the model extracts more and more features of the dataset. For instance, 4x4 just learns the form of a face, 8x8 learns to generate face and eyes, and so on.
As the resolution increases, more features are extracted like 1024 x 1024 extracts wrinkles, but the time taken to train 1024 x 1024 is in days. So, if we just stop the training at 512 x 512, the difference will not be much, but it will save days of training. This is an added advantage as the later resolution take more time combined than the previous resolutions.
Due to our limitation of training resources, we were able to experiment it on 64 x 64 to up sample it to 256 x 256 using SRGAN.

## VI. DATA AND RESULTS

The quality of the generated image is important to evaluate as we are upscaling the image by 4x without training. There are three factors that we have evaluated. Sliced Wasserstein Distance, Multi Scale Structural Similarity Index (MSSSIM), and Inception Score [17]. Table 2 summarizes our experiment.
The idea behind the sliced p-Wasserstein distance is to first obtain a family of marginal distributions (i.e., one dimensional distributions) for a higher-dimensional probability distribution through linear projections (via Radon transform), and then calculate the distance between two input distributions as a functional on the p-Wasserstein distance of their marginal distributions. In this sense, the distance is obtained by solving several one-dimensional optimal transport problems, which have closed-form solutions.
The structural similarity (SSIM) index measures perceived quality by quantifying the SSIM between an image and a reference image. This function calculates the MS-SSIM index by combining the SSIM index of several versions of the image at various scales. The MS-SSIM index can be more robust when compared to the SSIM index about variations in viewing conditions.
The IS takes a list of images and returns a single floating-point number, the score. The score is a measure of how realistic a GAN's output is. In the words of its authors, "we find (the IS) to correlate well with human evaluation (of image quality)". It is an automatic alternative to having humans grade the quality of images. The score measures two things simultaneously: The images have variety (e.g., each image is a different breed of dog). Each image distinctly looks like something (e.g., one image is clearly a Poodle, the next a great example of a French Bulldog)
If both things are true, the score will be high. If either or both are false, the score will be low. A higher score is better. It means your GAN can generate many different distinct images. The lowest score possible is zero. Mathematically the highest possible score is infinity, although in practice there will probably emerge a non-infinite ceiling.

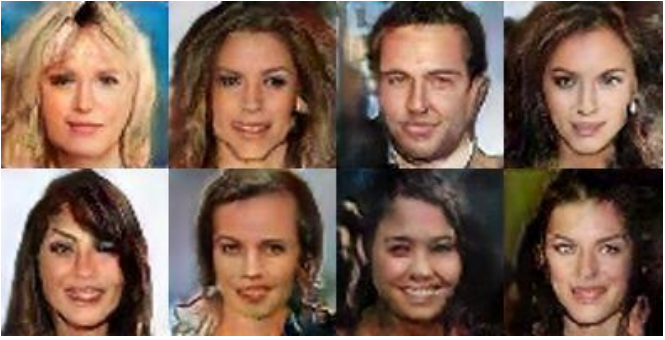

Fig. 3. The output of Pro GAN using Depthwise separable Convolution trained on 50 epochs on each resolution until 64 x 64 resolution.

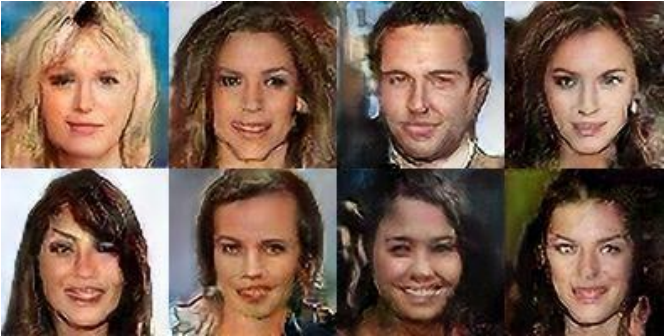

Fig. 4. The output after passing through the SRGAN model.

| SR No. | Metric | Progressive GAN | SRGAN |
|---|---|---|---|
| 1. | Sliced Wasserstein Distance | 240.65 | 381.86 |
| 2. | MSSSIM | 0.1592 | 0.1698 |
| 3. | Inception Score | 2.127 ± 0.263 | 2.138 ± 0.130 |

Table. 2. Evaluation of the pipeline

## VII. Conclusion

With the world moving towards faster and better ways of speeding up computation, non-traditional ways of reducing the complexity of heavy computation algorithms comes in handy. We proposed a new pipeline to reduce training of one of the important algorithms to augment high resolution dataset. Although we have proven our pipeline for low resolution, the training time graph depicted in Fig. 2, the original training time diverges with our method proposed and hence it will be more efficient for high resolution images. The SRGAN up sampling method is like an added advantage where we can just skip training of more than one steps (i.e., resolutions) for extraction of non-essential features like the wrinkles which are extracted in the last step and require day to train.

The future scope of this is to test the pipeline for the high resolution image which we were unable to implement in due to limited resources.

ACKNOWLEDGMENT

This research was supported by Vishwakarma Institute of Technology, Pune, India.